# GALACTIC COSMIC RAYS – CLOUDS EFFECT AND BIFURCATION MODEL OF THE EARTH GLOBAL CLIMATE.
# PART 2. COMPARISON OF THEORY WITH EXPERIMENT

[1]Vitaliy D. Rusov [a,b,c], Vladimir N. Vaschenko [c], Elena P. Linnik[a], Oksana T. Myhalus [a], Yuriy A. Bondartchuk[a], Vladimir P. Smolyar[a], Sergey I. Kosenko [a],
Strachimir Cht. Mavrodiev [d], Boyko I. Vachev [d]

[a] Odessa National Polytechnic University, Shevchenko av. 1, Odessa, 65044 Ukraine

[b] Bielefeld University, 25, University Str., Bielefeld, 33615 Germany

[c] Ukrainian National Antarctic Center, 16, Tarasa Schevchenko Blvd., Kiev, 01601 Ukraine

[d] Institute for Nuclear Research and Nuclear Energy, 72, Tsarigradsko shaussee blvd. 1784, Sofia, Bulgaria

## Abstract

The solution of energy-balance model of the Earth global climate and the EPICA Dome C and Vostok experimental data of the Earth surface palaeotemperature evolution over past 420 and 740 kyr are compared.

In the framework of proposed bifurcation model (i) the possible sharp warmings of the Dansgaard-Oeschger type during the last glacial period due to stochastic resonance is theoretically argued; (ii) the concept of climatic sensitivity of water in the atmosphere, whose temperature instability has the form of so-called hysteresis loop, is proposed, and based of this concept the time series of global ice volume over the past 1000 kyr, which is in good agreement with the time series of $\delta^{18}O$ concentration in the sea sediments, is obtained; (iii) the so-called "$CO_2$ doubling" problem is discussed.

---

[1] Corresponding author: Prof. Rusov V.D., Head of Department of Theoretical and Experimental Nuclear Physics, Odessa National Polytechnic University, 1, Shevchenko ave., Odessa, 65044, Ukraine,

Tel.: +380 487 348 556 , Fax:+380 482 641 672, E-mail: siiis@te.net.ua

# 1. Introduction

As is shown in Part 1 of the present paper (Rusov et al, 2010), the basic equation of energy-balance model of the Earth global climate is the bifurcation equation (with respect to the Earth surface temperature) of assembly-type catastrophe with the two governing parameters, which describe the insolation variations and the Earth magnetic field variations (or the variations of cosmic ray intensity in the atmosphere). A general bifurcation problem of this energy-balance model (see Eqs. (20)-(23) and (26)-(28) in Ref. (Rusov et al, 2010)) consists in determination of the global temperature $T(t)$ and its increment $\Delta T(t)$ and is reduced to finding the stable solution set of equations:

$$\frac{\partial}{\partial T} U^*(T,t) = T_t^3 + a(t) \cdot T_t + b(t) = 0, \qquad (1)$$

where

$$a(t) = -\frac{1}{4\delta\sigma} a_\mu H_\oplus(t),$$

$$b(t) = -\frac{1}{4\delta\sigma}\left[\frac{\eta_\alpha S_0}{4} + \frac{1}{2}\beta + \frac{1}{2}b_\mu H_\oplus(t)\right]$$

and

$$\frac{\partial}{\partial T} \Delta U^*(\Delta T,t) \cong \Delta T_t^3 + \tilde{a}(t) \cdot \Delta T_t + \tilde{b}(t) = 0, \qquad (2)$$

where

$$\tilde{a}(t) = -\frac{37.6}{\sigma T_t^3} a_\mu H_\oplus(t) = -\tilde{a}_0 H_\oplus(t);$$

$$\tilde{b}(t) = -\frac{37.6}{\sigma T_t^3}\left[\eta_\alpha \frac{S_0 + \Delta\hat{W}(t)\sigma_S}{4} - 4\delta\sigma T_t^3 + \frac{1}{2}\beta + \frac{1}{2}(2a_\mu T_t + b_\mu)H_\oplus(t)\right] =$$

$$= -\tilde{b}_0\left[\eta_\alpha W_{reduced}(t) - 4\delta\sigma T_t^3 + \frac{1}{2}\beta + \frac{1}{2}(2a_\mu T_t + b_\mu)H_\oplus(t)\right];$$

$U^*(T, t)$ describes with an accuracy up to constant the so-called "inertial" power of heat variations in the ECS; $U^*(T, t)$ is the variation of $U^*(T, t)$, $H_\oplus$ is the intensity of terrestrial magnetism, $T_t$ is the average global temperature of the Erth surface at the time $t$, K; $\Delta T_t$ is the variation of $T_t$ ; $S_0$=1366.2 Wm$^{-2}$ is "solar constant" (Frohlich and Lean 1998); $\delta$=0,95 is coefficient of gray chromaticity of the Earth surface radiation; $\sigma=5{,}67\cdot 10^{-8}$ is the Stephen-Boltzmann constant, $Wm^{-2}K^{-4}$; $\eta_\alpha=0{,}0295\ K^{-1}$ (see Eq. (19) in Ref. (Rusov et al., 2010)); $\beta$ is the accumulation rate of carbon dioxide in the atmosphere normalized by unit of temperature, $kgK^{-1}$; $a_\mu$ и $b_\mu$ are constants, whose dimensions are $Wm^{-2}K^{-2}$ and $Wm^{-2}K^{-1}$, respetively; $\Delta\hat{W}(t)$ is the insolation reduced normalized variation; $\sigma_S$ is the roof-mean-square deviation; $4W_{reduced}=S_0+\Delta\hat{W}(t)\sigma_S$ is the reduced annual insolation.

A purpose of this paper is a comparative analysis of solution of Eqs.(1)-(2) of the energy-balance model of the Earth global climate (Rusov et al, 2008) and well-known experimental time series of palaeotemperature (e.g., the Vostok ice core data (Petit et al, 1999) over the past 420 kyr and the EPICA ice core data (EPICA community members, 2004) over the past 740 kyr.

## 2. Hypothesis of the Northern hemisphere substrate and ice sheet response function

Before to search the solution of the system of the initial Eq. (1) and the "perturbed" Eq.(2) (with an allowance for initial and boundary conditions), we shall do a few important remarks concerning the specific character of Eq. (2) perturbation.

It is known that "ice sheets influence climate because they are among the largest topographic features on the planet, create some of the largest regional anomalies in albedo and radiation balance, and represent the largest readily exchangeable reservoir of fresh water on Earth" (Clark et al, 1999). Moreover, recently influence of the Northern hemisphere ice sheets on the global climate change is discussed within the framework of a hypothesis of the so-called Northern substrate effect, which explains the mechanisms that nearly synchronize the climate of the Northern and Southern Hemispheres at orbital time scales despite asynchronous insolation forcing in these Hemispheres. The essence of hypothesis consists in the fact that "the effect of the substrate underlying the Northern Hemisphere ice sheets is to modulate ice-sheet response to insolation forcing" (Clark et al, 1999).

In order to take into account the influence of the Northern substrate in our equations, it is necessary to consider such insolation variations, which influence directly the value of ice-sheet response (Clark et al, 1999). At the same time, the basic equation of state (1) comprises the average annual insolation in the form of the time-dependent parameter $W(t)\cong S_0/4(1-e^2)$, which weakly depends on the eccentricity $e(t)$. Therefore, the insolation variations can not be the reason for the non-linear response of ice sheets. On the other hand, the perturbed equation of state (2) contains the entropy term $\partial\Delta G_{w+v}/\partial(\Delta T)$, which describes by definition (see Eqs. (16) and (17) in Ref. (Rusov et al, 2010)) the mass increment of water and water vapour $\Delta m_{w+v}$ in the atmosphere in the following way:

$$\frac{\partial}{\partial(\Delta T)}\Delta G_{w+v}(\Delta T,t) = \left(\tilde{a}_0\Delta T + \tilde{b}_0\frac{2a_\mu T_t + b_\mu}{2}\right)H_\oplus(t) \sim \Delta m_{w+v}, \qquad (3)$$

where $\Delta m_{w+v} = \langle\rho\rangle(a_{wv}+b_{wv}\Delta T)H_\oplus(t)$; $\langle\rho\rangle$ is the averaged specific gravity of combined water (water vapour and liquid water ) in the atmosphere. Recall that $H_\oplus = M_t\chi_{t=0}/M_{t=0}\chi_t$ (see Eq.(15) in (Rusov et al, 2010)). Necessary experimental data of the magnetization $M_t$ and magnetic susceptibility $\chi_t$ at millennial time scales for the time $t \in [0, 2.25]$ million years are obtained by Yamasaki and Oda (2002).

It is obvious, that the variations of additional mass of water and water vapour in the atmosphere $\Delta m_{w+v}$ within the framework of our model are, in essence, the variations of the evaporated fresh water on the Earth. In this context, there are the two reasons, which predetermine the further way of Eq. (2) perturbation. On the one hand, according to the Northern substrate hypothesis just the Northern Hemisphere ice volume is reservoir of fresh water on the Earth, which plays one of the key roles in the mechanism of vigorous ice-sheet response to insolation forcing. On the other hand, as is known, the vicinity of the 65$^o$ Northern and Southern latitude are the places where the cosmic rays are of high flux. At the same time, ice formed in latitude 65$^o$N has a significantly longer "lifetime" than ice at same latitude in the Southern hemisphere. This takes place because it usually forms on a land surface in the Northern hemisphere and not on a water surface of southern oceans, which, moreover, are heated by global warm stream at these latitudes (~65$^o$ S). If to take into account that insolation variations are maximal just in vicinity of 65$^o$ northern and southern latitudes, it becomes clear, why continental ice formed near 65°N plays the role of a super amplifier of insolation variations in the region of 65° N, whereas calved ice in the region of 65$^o$ S has the weak response to high insolation forcing due to its not great volume. Apparently, just for this reason the values of long-term midmonth insolation variations at 65°N for July are used in the simulation of the Northern Hemisphere ice volume (Berger and Loutre, 2002).

Therefore, to emulate the Northern substrate effect in our model, we "perturb" the average annual insolation $\langle W_0\rangle \cong S_0/4$ in Eq. (2) so that the variations of water and water vapour mass in the atmosphere $\Delta m_{w+v}$ are modulated not only by the Earth magnetic field variations (see Eq. (3)) but also by midmonth insolation variations at 65$^o$N for July (by way of temperature variations).

With that end in view, we write down a balance equation (similar to Eq.(25) in Ref. (Rusov et al, 2010)) but for the sum of the normalized variations of insolation $\Delta W_i$ over all $i$-th latitudes at the time $j$:

$$\frac{1}{\gamma}\sum_{i=-90^\circ}^{+90^\circ}\gamma_i W_{ij} = \frac{1}{\gamma}\sum_{i=-90^\circ}^{+90^\circ}\gamma_i\langle W_{ij}\rangle + \frac{1}{\gamma}\sum_{i=-90^\circ}^{+90^\circ}\gamma_i\Delta W_{ij}\sigma_{ij}, \qquad (4)$$

$$\frac{1}{t_n}\sum_{j=0}^{t_n}W_{ij} = \langle W_{ij}\rangle,$$

where $\gamma_i/\gamma$ is the part of the surface corresponding to *i*-th latitude; $-90° \leq i \leq 90°$; $t_n$ is the time interval of averaging; $\sigma_{ij}$ is standard deviation of the current value of insolation $W_{ij}$ from the average $<W_{ij}>$ at *i*-th latitude at the time $j \in [0, t_n]$. Note that the partition number of the sampling *n* is usually equal to 100 at $t_n=1000$ kyr.

Now we introduce the so-called response function of continental and calved ice $h(i)$, which describes ice-sheet response to insolation forcing (Clark et al, 1999). As the ice volume at latitude 65° S (the Antarctic peninsula) can be neglected, the response function $h(i)$ looks like

$$h(i) = \frac{\gamma}{\gamma_{65}} \frac{\langle W_0 \rangle}{\langle W_{65} \rangle} \delta(i-65°) + \sum_{k=64°}^{+90°} \frac{\gamma}{\gamma_k} \frac{\langle W_0 \rangle}{\langle W_k \rangle} \delta(i-k) + \sum_{k=-64°}^{-90°} \frac{\gamma}{\gamma_k} \frac{\langle W_0 \rangle}{\langle W_k \rangle} \delta(i-k), \quad (5)$$

where $-90° \leq i \leq +90°$ and

$$\langle W_0 \rangle = \left\langle \frac{S_0}{4(1-e^2)} \right\rangle \cong \frac{S_0}{4} \quad (6)$$

is by definition the average annual insolation weakly dependent on the eccentricity $e(t)$ of the Earth elliptic orbit.

It is clear, that introduced in this manner Eq. (5) emulates the anomalous behaviour of the response function of continental and calved ice of the Earth. Then "perturbation" of the average annual insolation $<W_0>$ caused by the response function (5) can be obtained by multiplication of the function (5) by Eq.(4). Moreover, if take into account that

$$\Delta W_i \sigma_i \cong -\Delta W_{-i} \sigma_{-i}, \quad (7)$$

the effect of response function (5) to Eq. (4) looks like

$$W_{65} \frac{\langle W_0 \rangle}{\langle W_{65} \rangle} = \langle W_0 \rangle + \frac{\langle W_0 \rangle}{\langle W_{65} \rangle} \Delta W_{65} \sigma_{65}. \quad (8)$$

Apparently, this is the formal mathematic presentation of the physical essence of the so-called Northern substrate (Clark et al, 1999).

The expression for the reduced average annual insolation (8) we rewrite in a more simple and convenient form

$$4W_{reduced}(t) = S_0 + \Delta \hat{W} \sigma_S, \quad (9)$$

where

$$W_{reduced}(t) \equiv W_{65} \frac{\langle W_0 \rangle}{\langle W_{65} \rangle}, \quad \langle W_0 \rangle \cong \frac{S_0}{4}, \quad \Delta \hat{W} \equiv \Delta W_{65}, \quad \sigma_S \equiv \frac{4\langle W_0 \rangle}{\langle W_{65} \rangle} \sigma_{65}. \quad (10)$$

At the same time the reduced standard deviation looks like

$$\sigma_S = \frac{4\langle W_0 \rangle}{\langle W_{65} \rangle} \sigma_{65} = \frac{4\langle W_0 \rangle}{k \langle W_{65}^L \rangle} \sigma_{65} = 53.45, \quad (11)$$

where $<W_0> = S_0/4$ is the average annual insolation; as the values of average insolation $<W_{65}> = k\langle W_{65}^L \rangle = k$ 437.457 Wm$^{-2}$) and standard deviation $\sigma_{65}=17.321$ at latitude $65^0$ N we use the calculated data $\langle W_{65}^L \rangle$ at $k=S_0/S_0^L=1366.2/1350=1.012$ of Laskar et al. (1993).

In general, to obtain the universal response function of the Earth surface it is necessary to take into account the full information of ice (fresh water) space-time distribution on the Earth surface. However, the computational experiment considered below shows that in a first approximation it is sufficient to use the Earth surface response function in the form (5).

### 3. Computational experiment

It is obvious, to solve bifurcation Eqs. (1) and (2), which describe the extreme temperature $T$ and its increment $\Delta T$, it is necessary to determine only the values of the three climatic constants $a_\mu$, $b_\mu$ and $\beta$ in Eqs. (27)-(28) from Ref. (Rusov et al, 2010) since the values of other parameters are usually known.

The value of $\beta$ we determine using the known estimation of heat energy power reradiated by carbon dioxide, i.e. $W_{CO_2} \sim 1.7$ Wm$^{-2}$ (Wigley, 2005). In our case this magnitude can be presented as the sum of reradiated energy variations $\Delta G_{CO_2}(T_{t=0})$ in Eq. (1) and $\Delta G_{CO_2}(\Delta T_{t=0})$ in Eq.(2) which in a linear approximation obeys the following equality

$$\frac{1}{\gamma}\left[\Delta G_{CO_2}(T_{t=0}) + \Delta G_{CO_2}(\Delta T_{t=0})\right] = \frac{1}{2}\beta T_0 = \frac{1}{2}W_{CO_2} = \frac{1}{2} \cdot 1.7 \quad \left[Wm^{-2}\right], \qquad (12)$$

where $H_\oplus(t=0)=1$; the coefficient of 1/2 implies the reemission isotropy of carbon dioxide. the As the initial condition we use below the value of the modern average global temperature $T_0$ (Ghil, 1980):

$$T_0 = T_{t=0} + \Delta T_{t=0} = 288.6 \pm 1.0 \quad [K]. \qquad (13)$$

Taking into account Eqs. (1) and (13), we have from Eq.(12)

$$\beta = 1.7/T_0 = 0.006 \quad \left[Wm^{-2}K\right]. \qquad (14)$$

The solution of the system of the initial Eq.(1) and "perturbed" Eq. (2) at the initial conditions (13) and (14) gives the following values of climatic constants

$$a_\mu = 0.222, \quad b_\mu = -127.249 \qquad (15)$$

and the final values of initial parameters of the system of the equations (1) and (2):

$$T_{t=0} = 286.031 \quad [K], \quad \Delta T_{t=0} = 2.100 \quad [K], \quad \beta = 0.006 \quad \left[Wm^{-2}K\right], \qquad (16)$$

which satisfy the degree of uncertainty of the modern global temperature $T_0$ described by Eq. (13).

Note, that analysis of the parameterized solution of the system of the equations (1)-(2) (i.e., the temperatures $T_t+\Delta T_t$) shows the high solution stability with respect to the relatively small

"perturbations" of the initial and boundary parameters (15)-(16). In other words, the small variations of climatic constants $a_\mu$ and $b_\mu$, lead to the small changes of the initial parameters $T_{t=0}$, $\Delta T_{t=0}$ and $\beta$. Anticipating things, it is important to note that the general solution of the equation system (1)-(2) or, more exactly, the time distribution of the theoretical palaeotemperatures $T_t+\Delta T_t$, according to the computational experiment practically does not change its form at the small variations of the climatic constants (15) and initial parameters (16). This means that the approximate invariance of the shape of time distribution of the theoretical temperature $T_t+\Delta T_t$ is predetermined by some attractor in the phase portrait of the physical system (1)-(2) and requires an individual consideration, but this is already another problem, going beyond the scope of the present paper.

Let us consider now the solution of Eq. (1) with respect to the temperature $T_t$ obtained with allowance for the parameters (15)-(16). First, the dependence of temperature $T_t$ on the relative magnetic field $H_\oplus$ (Fig. 1) is considered.

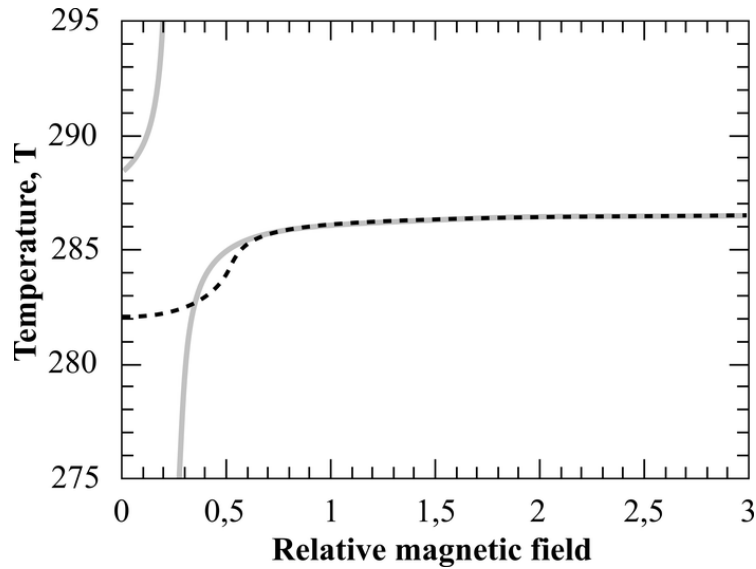

Fig. 1. Theoretical dependence of the temperature $T_t$ on the relative magnetic field $H_\oplus$ (solid curve) obtained from Eq.(1) (see Eq. (29) in Ref. (Rusov et al, 2010)) with allowance for Eqs.(21)-(22) from Ref. (Rusov et al, 2010). In computations the dependence, which has hysteresis shape (dotted curve) and meets the condition (17), is used. Explanation is in text.

As follows from Fig. 1, the temperature $T_t$ is very low when $H_\oplus \leq 0.5$. The existence of such evidently non-physical range of the temperatures $T_t$ is a consequence of rough approximation used in deriving Eq. (14) in Ref. (Rusov et al, 2010). In the view of physics, it, apparently, means that this equation does not take into consideration a threshold nature of solar wind influence on the galactic cosmic ray flux and Earth magnetosphere by reconnection. In other words, Eq. (14) does not take into consideration such a minimum value of solar wind magnetic induction when the intensity of galactic cosmic ray "sweeping-out" and interaction of solar wind with the Earth magnetosphere though are "minimal" but nonzero. Moreover, this low interaction between solar wind and the Earth magnetosphere is veiled by the intrinsic variations of the Earth's magnetic field due to the specificities of the Earth liquid core kinetics and the stochastic properties of magnetic field energy source, i.e. slow nuclear burning on the boundary of liquid and solid phases of the Earths core (the georeactor of 30 TW) (Rusov et al, 2007).

On the other hand, it is clear that the dependence of the temperature $T_t$ on the value of relative magnetic field $H_\oplus$ (Fig. 1) by virtue of Eq.(1) type has to be of a hysteresis type. For this reason, in the range $H_\oplus \leq 0.5$ the real curve in Fig. 1 must to be a several Kelvins lower (for example, 3-5 K) than the part of curve located in the range $H_\oplus \geq 0.5$. The subsequent computational experiments have shown that of all solutions (i.e., the temperatures $T_t+\Delta T_t$) of the equation system (1)-(2) at $H_\oplus \leq 0.5$ the "reference" temperature $T_{H_\oplus \leq 0.5}$ of the Earth climatic system (ECS) (Fig. 1)

$$T_{H_\oplus \leq 0.5} = 282 \quad [K], \quad when \quad H_\oplus \leq 0.5 \qquad (17)$$

is the best value for experimental data fitting (Petit et al, 1999; EPICA community members, 2004).

Fig. 2d shows the plotted computational solution of Eqs. (1)-(2) which describes time evolution of temperature $T_t+\Delta T_t$ with respect to the average temperature $T_{t=0}$=286 K. The experimental data of palaeotemperature over the past 420 kyr and 740 kyr obtained from the Vostok ice core (Petit et al, 1999) and the EPICA ice core (EPICA community members, 2004), respectively, are shown for comparison in Fig. 2e.

Thus, the good agreement of experimental (Fig. 2e) and theoretical (Fig. 2d) data is the validation of main assumption in our model (recall we assume that the ECS temperature is determined by both the insolation variations and GCR intensity variations or, equivalently, by the Earth magnetic field variations). The last supposition is legitimate because the GCR flux in a heliosphere is stably constant at time scales less than 1000 kyr.

At the same time, it is necessary to pay attention to so-called anomalous palaeotemperature jumps. These temperature jumps are marked with red vertical lines in Fig. 2e. By anomalous palaeotemperature jumps we understand sharp temperature rise, which, as a rule, takes place at once after synchronous passing of the insolation (see Fig. 2b) and the Earth relative magnetic field (see Fig. 2c) through a minimum. Below we attempt to find the way for solving such a physical problem and thereby to explain, why such jumps are not described still within the framework of the proposed bifurcation model of global climate.

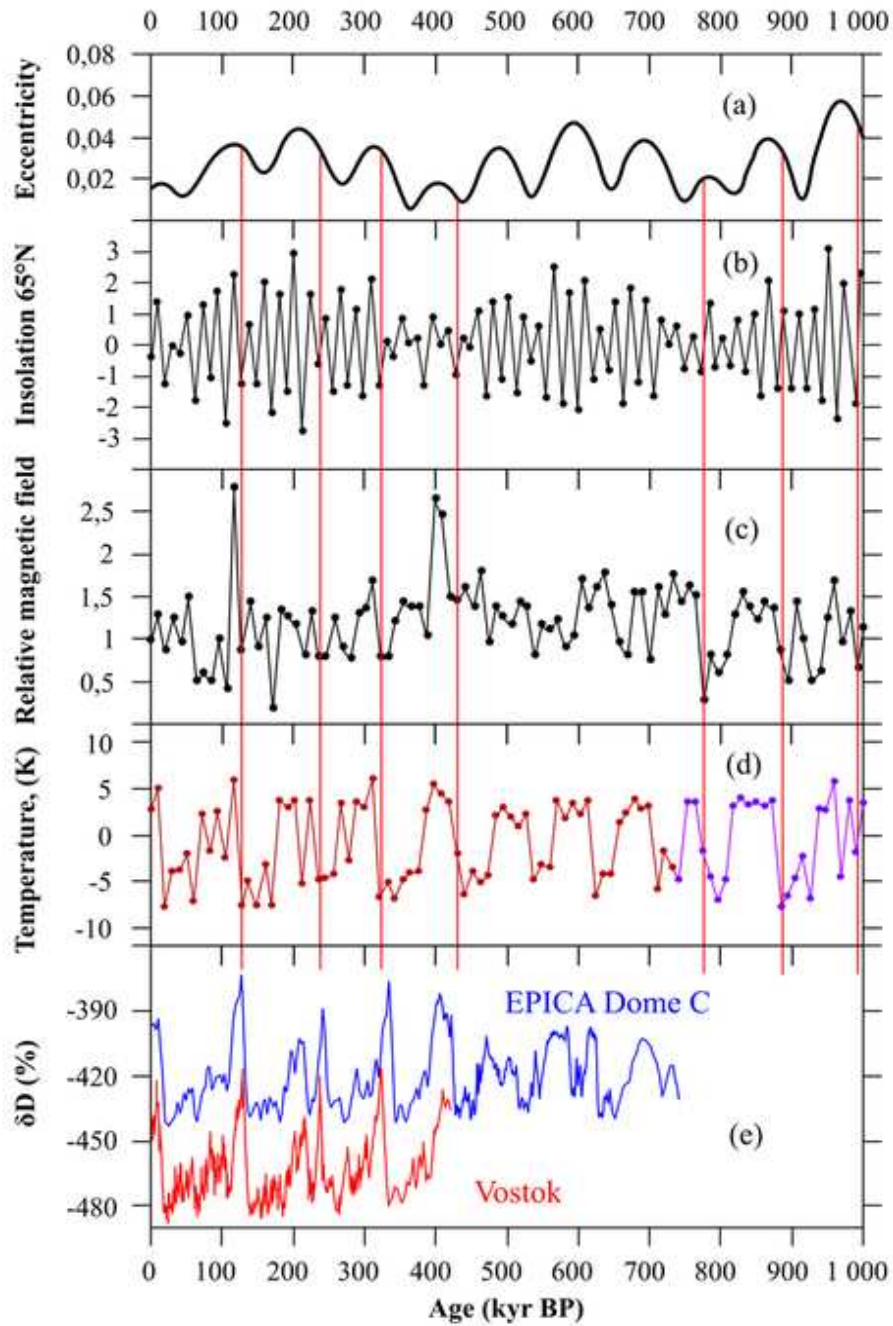

Fig. 2. The model of the ECS temperature response to insolation variations and Earth magnetic field variations in the 1000 kyr time interval. In the figure is shown the time evolution of (a) the Earth orbital eccentricity, (b) the insolation variations in latitude $65°N$ (summer) (Laskar et al, 1993), (*c*) the variations of the Earth magnetic paleointensity calculated on the basis of Ref. (Yamasaki and Oda, 2002), (d) the variations of the Earth surface model temperature (with respect to the average current temperature $T_{t=0}=286.0$ K) obtained by solving the system of equations (1)-(2), (e) the variations of deuterium concentration $\delta D$ (the conditional analogue of paleotemperature) measured in ice cores within the framework of the Antarctic projects "Vostok" over the past 420 kyr (Petit et al, 1999) and EPICA Dom C over the past 740 kyr (EPICA community members, 2004). The anomalous jumps of variations of deuterium isotopic concentration $\delta D$ are indicated by arrows. The vertical red lines indicate the time of possible anomalous temperature jumps and variations of deuterium isotopic concentration $\delta D$, respectively.

Comparing the time series of the insolation, relative magnetic field and experimental palaeotemperature (Fig. 2) it is possible to conclude that neither insolation nor the Earth magnetic field can be such an energy source, which can generate these temperature anomalies. By virtue of the energy conservation law, an additional pulsed heat source must to exist in the energy balance of the ECS (see Eq. (9) and Fig. 5 (Rusov et al, 2010)) in order to explain this heat paradox. Let us consider the physical mechanism, which, in our opinion, predetermines the operation of such a source.

It is known that the magnetic field intensity of solar wind affecting on the Earth magnetosphere at millennial time scales is modulated by the eccentricity of Earth elliptic orbit. When the changes of the interplanetary magnetic field induced by solar wind exceed a certain threshold, by virtue of the Faraday law of electromagnetic induction they can become the reason for origin of inductive currents in the Earth liquid core. In other words, when the eccentricity verges toward its local maximum the magnetic flux variations of interplanetary field can cause the considerable variations of the Earth magnetic field, which, in its turn, are capable to induce electric current in the Earth liquid core. Such an additional current, which by virtue of the Le Chatelier law is oppositely directed relative to direction of convection current in the Earth liquid core, can partly disable the convection responsible for the Earth magnetic field generation. The convection partial blocking, on the one hand, leads to abrupt decrease in the rotation velocity of the Earth liquid core and, on the other hand, decreases heat removal from the separation surface of the liquid and solid phases of the Earth core. This can result in the significant increase in temperature in this region, where, as is supposed, the georeactor of 30 TW is operating (Rusov et al, 2007).

Now let us consider possible consequences. At first, we consider the consequences for the ECS due to the sharp decrease of the Earth liquid core rotation velocity. By virtue of the angular moment conservation law, the angular velocities of the mantle $\omega_{mantle}$, liquid $\omega_{liq}$ and solid $\omega_{solid}$ phases of the Earth are connected by the following relation

$$I_m \omega_{mantle} + I_l \omega_{liquid} + I_s \omega_{solid} = const, \qquad (18)$$

where $I_m$, $I_l$ and $I_s$ are the inertial moments of the mantle, liquid and solid phases of the Earth core, respectively.

It is obvious that by virtue of Eq. (18) the decrease in angular velocity of the Earth liquid core leads to the increase not only of angular velocity of the Earth solid core but also of the mantle rotation velocity. This means that the sharp change of mantle rotation velocity will cause strong friction between the lithosphere and surface layer of the atmosphere that, in its turn, will lead to the sharp increase in average temperature of the Earth surface and troposphere. Thus, such a physical mechanism of solar-terrestrial relation, in our opinion, can be responsible for sufficiently infrequent (approximately once a 100 kyr) but powerful heat pulses (Fig. 2e) discharged into the atmosphere from the Earth surface.

Now let us consider geophysical consequences induced by the significant increase of the Earth liquid core temperature. It is known that the induced fission cross-section of $^{238}$U, which is the natural fuel for the georeactor (Rusov et al, 2007), strongly depends on the ambient temperature $T$ ($\sigma_f \sim T^4$) in the resonance region. It means that the nuclear burning rate on the boundary of the liquid and solid phases of the Earth core increases with ambient temperature growth. Moreover, the nuclear burning rate will continue increase until the temperature on the separation surface of the liquid and solid phases of the Earth core will become significantly higher than the average temperature of liquid core that, in its turn, will lead to the recovery of the ordered convection in the liquid core. At the same time, direction of the recovered convection can coincide or not coincide with the original direction of convection in the liquid core prior to its distrurbance. In the case when directions of the recovered and disturbed convection do not coincide, a very interesting effect is observed, i.e. so-called the Earth magnetic field inversion. Obviously, that the average periodicity of such inversions must correlate with period of the Earth orbit eccentricity. This agrees with conclusions of Consolini and De Michelis (2003), who have researched the stochastic mechanism of geomagnetic poles inversions. So, in case of abrupt increse of the Earth liquid core temperature, the georeactor plays the role of an energy stabilizer of ordered convection in the Earth liquid core and the Earth magnetic field, respectively.

## 4. Discussion and conclusions

To analyze the behaviour of a complex climatic systems, it is convenient to use the notion of so-called climatic sensitivity, e.g. from the standpoint of definition proposed by Wigley and Raper (2005). In our case the climatic sensitivity $\lambda_{w+v}$ with consideration of Eq.(3) (see Eq. (31) in Part 1 (Rusov et al, 2010)) describes the of change of the total water mass $\Delta m_{w+v}$ (water vapour and liquid water) in the atmosphere with respect to the temperature increment $\Delta T$:

$$\lambda_{w+v} = -\frac{1}{\Delta t}\frac{\partial}{\partial(\Delta T)}\Delta G_{w+v}(\Delta T, t) \sim \frac{1}{\Delta t}\Delta m_{w+v}, \qquad (19)$$

where $\Delta t = 10$ Kyr is time scale resolution (see Fig. 2).

The physical meaning of this value becomes clear from Fig.3, where the climatic sensitivity sufficiently discrubes the temperature bistability intrinsic to the "potential" of assembly-type catastrophe (see Eqs.(20) and (26) in Ref.(Rusov et al, 2010)). Also, the decrease (increase) of increment rate of total water mass (vapour and liquid) in the atmosphere $\lambda_{w+v}$ predetermines, by virtue of the conservation law, the same increase (decrease) of increment rate of total fresh water mass $\Delta F_{fwf}$ in the ECS. This shows that there is a direct interrelation between the global temperature of the ECS and the rate of fresh water mass change (Fig.2) whose evolution is provided by the global ocean circulation (Wunsch, 2002). Moreover, such an interrelation can also display the temperature instability in the form of hysteresis loop analogous to that shown in Fig.3.

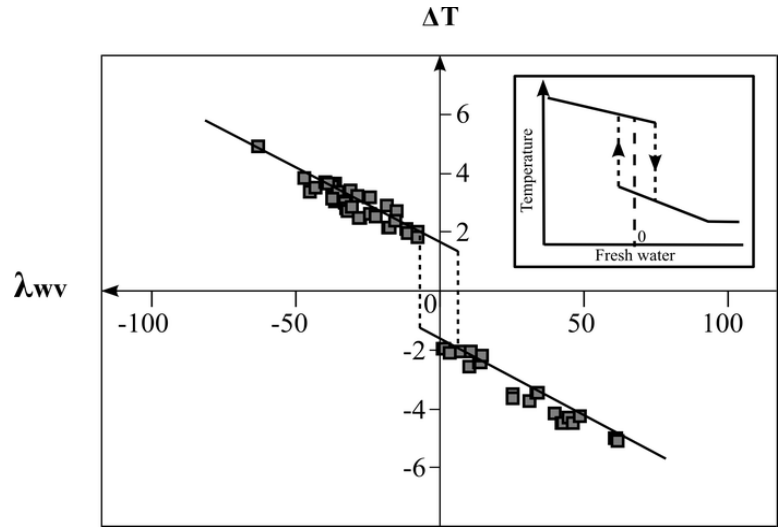

Fig. 3. The dependence of variations of the theoretical increment of temperature (■) on the climatic sensitivity $\lambda_{w+v}$ calculated by Eqs. (19) and (3). Insert – the adapted schematic sketch of known hysteresis dependence of temperature on the rate of increment of fresh water total mass in high-latitude experiments in the North Atlantic of Ganopolski and Rahmstorf (2001).

Thus, along with the parameter $\lambda_{w+v}$, the parameter $\Delta F_{fwff}$, which determines the rate of change of fresh water mass with respect to the global temperature, is also the climate sensitive parameter. In general, it is possible be suppose that all physical, chemical, and biological parameters, which display the temperature instability in the form of the hysteresis loop, belong to a class of variable climate-sensitive parameters, but in the general case at different geological time-scales. Therefore, the climate-sensitive parameters, which belong to one and the same geological time-scale, must be arcwise connected between themselves. This can be easily shown (Fig.4) by the example of the following series of arcwise-connected parameters:

$$-\lambda_{w+v} \sim \Delta F_{fwf} \sim I_V \,, \qquad (20)$$

where $I_V$ is the global ice volume.

Moreover, the apparent correlation (see Fig. 4) between the known time series of $\delta^{18}O$ record from marine sediments (a proxy for global ice volume) (Bassinot et al, 1994; Tidemann et al, 1994) and similar time series of climatic sensitivity $\lambda_{w+v}$ calculated by Eqs. (19) and (3) indicates both the existence of direct linear connection between these parameters and predictive force of the discussed bifurcation model of the Earth global climate (see Figs. 2 and 4).

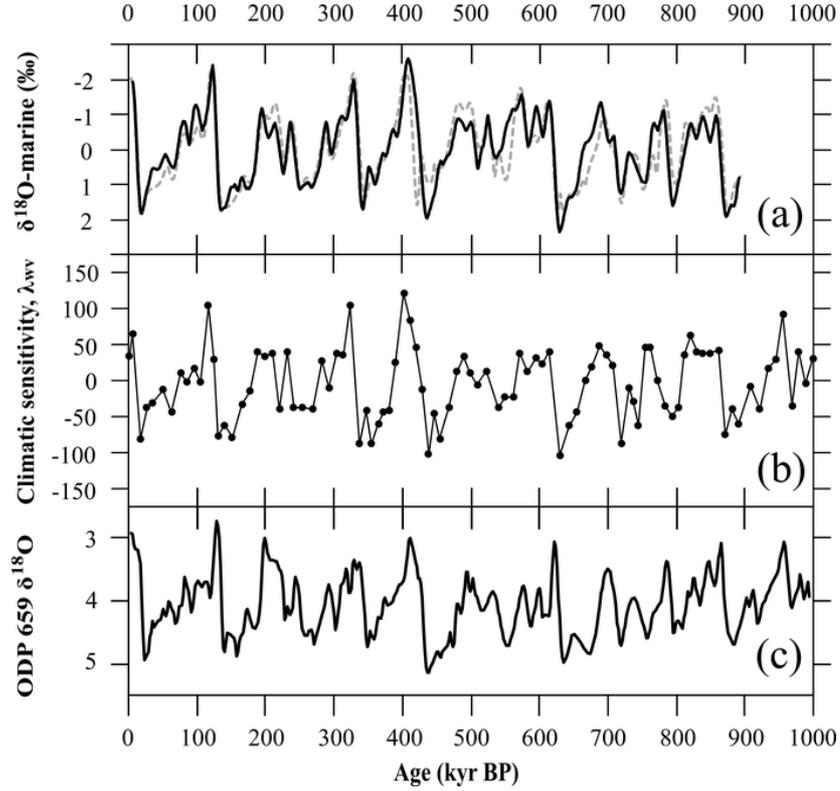

Fig. 4. Comparison of the theoretical time series of climatic sensitivity $\lambda_{w+v}$ calculated by Eqs. (3) and (19) (b) with the time series of $\delta^{18}O$ isotopic concentration (the conditional analogue of ice volume) measured in the deep-water experiments: (a) Bassinot et al. (1994) (solid blue line) and Imbrie at al. (1993) (dashed red line), (c) Tidemann et al. (1994).

On the other hand, it is known that the temperature instability of global climate like other analogous instabilities of wide class of bistable systems can cause the stochastic resonance under certain conditions (Benzi, 2007; Gamaitoni et al, 1998; Kramers, 1940; Hanggi et al, 1990; Moss, 1994). For example, using an ocean-atmosphere climate model (which belongs to bistable models), Ganopolski and Rahmstorf (2002) ascertained that the stochastic resonance is the main mechanism (of climate variability at secular time scale) during last glacial period. Moreover, the simulated warm events are in good agreement with the properties of so called the Dansgaard-Oeschger events recorded by isotope measurements of $\delta^{18}O$ in the Greenland ice core (Ganopolski and Rahmstorf, 2002; Braun et al, 2005; Andersen et al, 2004). This is an important outcome, which, moreover, is in good agreement with other experimental observations. Because it would be interesting to show how this outcome is easily described by the proposed bifurcation model of the Earth climate.

The stochastic resonance can be obtained, if the stochastic system (for example, the ECS) by varying the noise intensity is tuned so that the modulated signal has a maximal multiplication (Benzi, 2007; Gamaitoni et al, 1998). The systems, in which the stochastic resonance is observed, are attributed to nonlinear systems with the characteristic dynamics, which, in some sense, is intermediate between the regular and chaotic dynamics. In the simplified form, the mechanism of stochastic resonance can be explained in terms of the Brownian particle motion in the bistable potential field, for example, under a white noise and simultaneously weak periodic signal.

In our case, the Earth climate system, which is in the asymmetric bistable potential field (see Eq.(26) in Ref. (Rusov et al, 2010)) can be considered as the Brownian particle. The "glacial" time interval, in which an appearance of the Dansgaard-Oeschge warm events can be expected, is in the 20 to 75 kyr range. On the one hand, this theoretical range coincides with the range of the Dansgaard-Oeschge actual events obtained in the NGRIP-experiments by the high-resolution ($\Delta t$=50 year) method of $\delta^{18}O$ concentration measurement, which is the effective indicator of global ice volume (Andersen et al, 2004). On the other hand, assuming that the coefficients of bistable potential field are constant in this range, Eq.(26) from Ref. (Rusov et al, 2010) can be rewritten in the form

$$\Delta U(\Delta T, t) = \frac{1}{4}\Delta T^4 - \frac{1}{2}\tilde{a}_{DO}\Delta T^2 - \tilde{b}_{DO}\Delta T, \quad \tilde{a}_{DO}, \tilde{b}_{DO} > 0. \tag{21}$$

As is shown in Ref. (Braun et al, 2005) the insolation $S(t)$ on secular time scale is under combined action of the two harmonic oscillations with the amplitudes $A_1$ and $A_2$ induced by so-called DeVries-Suess (Wagner et al, 2001) and Gleisberg (Peristykh and Damon, 2003) solar cycles with the periods $T_1$= 210 years and $T_2$= 86.5 years:

$$S(t) = S_0 + \Omega(t) = S_0 + A_1 \sin(\omega_1 t + \varphi_1) + A_2 \sin(\omega_2 t + \varphi_2), \tag{22}$$

where $\omega_1=2\pi/T$ and $\omega_2=2\pi/T_2$ are frequencies of this cycles.

As is experimentally and theoretically shown (Ganopolski and Rahmstorf, 2002; Braun et al, 2005), superposition of these two frequencies generates the resulting oscillation with period of 1470 years. Recall that in this case the term corresponding to insolation in Eq. (28) from Ref. (Rusov et al, 2010) at the secular time scale has the form:

$$4W_{reduced}(t) = S_0 + \Omega(t) + \Delta \hat{W}(t)\sigma_S. \tag{23}$$

On the other hand, the high time resolution ($\Delta t$=50 years) in the NGRIP-experiments and the known behavior of temperature spectral density of the ECS (Kutzbach and Bryson, 1974) make it possible to conclude that the combined action of white noise and 1/$f$- noise is the main reason of the ECS temperature variations in this time interval. Then for the over-damped system (in whose equation of motion the inertia term $m\ddot{x}$ as compared to the friction $\dot{x}$ may be neglected) under action of weak harmonic signal, for example, of the $A_1\sin(\omega_1 t+\varphi_1)+ +A_2\sin(\omega_2 t+\varphi_2)$ type, the dependence $\Delta T(t)$ can be obtained on the basis of the Langevin type equation

$$\dot{x} = -\partial(\Delta U^*)_x(x,t) + \zeta(t), \quad x(t) \equiv \Delta T(t) \tag{24}$$

or with allowance for and Eq. (23) and also Eqs. (26)-(28) from Ref. (Rusov et al, 2010)

$$\dot{x} = \tilde{a}_{DO}x - x^3 + \tilde{b}_{DO} + \tilde{b}_0\eta_\alpha[A_1\sin(\omega_1 t + \varphi_1) + A_2\sin(\omega_2 t + \varphi_2)] + \zeta(t), \tag{25}$$

where $\dot{x}$ is the friction force; $\zeta(t) = \xi(t) + \eta(t)$ is the additive mixture of white noise and $1/f$-noise, respectively. When the noise is high the term $\tilde{b}_{DO}$ causing an asymmetry of potential in Eqs. (21) and (25) can be neglected.

In scientific literature the analogue of this equation (25) is the well studied canonical equation, which describes the stochastic resonance for the Brownian particle motion in symmetric bistable potential field. This equation with our designations has the following form

$$\dot{x} = \tilde{a}_{DO}x - x^3 + A\sin\omega t + \xi(t), \quad x(t) \equiv \Delta T(t), \tag{26}$$

where the restriction $Ax_m \ll \Delta(\Delta U^*(x,t)) = \Delta U^*(0) - \Delta U^*(x_m)$ is imposed on weak periodic signal amplitude; $\pm x_m = (\tilde{a}_{DO})^{1/2}$ is the coordinate of potential minimum $\Delta U^*$; $\Delta(\Delta U^*(x,t)) = \tilde{a}_{DO}^2/4$ is the potential barrier height; $\xi(t)$ is the white noise of Gaussian type, which has the zero mean $\langle\xi(t)\rangle=0$ and is described by the correlation function of the form $\langle\xi(t), \xi(0)\rangle = 2D\delta(t)\rangle$.

In most cases to identify the stochastic resonance it is convenient to use the experimental distribution of particle residence time in the potential well (Gammatoni et al, 1995; 1998)

$$P(t) \sim \exp(-W_K t), \tag{27}$$

$$W_K = \frac{1}{\tau_K} = \frac{1}{2\pi}\left\{\left|\frac{d^2(\Delta U^*)}{dx^2}\right|_{x=0}\left|\frac{d^2(\Delta U^*)}{dx^2}\right|_{x=x_m}\right\}^{1/2}\exp\left(-\frac{\Delta(\Delta U^*) \pm x_m A\sin\omega t}{D}\right), \tag{28}$$

where $W_K$ and $\tau_K$ are the modified "Kramers" rate and time, respectively (Kramers, 1940; Hanggi et al, 1990; Moss, 1994), which describe the rate and time of the transition of excited system through a potential barrier under the conditions of weak periodic perturbation.

Because of the periodic component in Eq. (28) the series of peaks corresponding to odd harmonics, which are centred on times multiple of the control signal half-period

$$T_n \cong (2n+1)\frac{\pi}{\omega} = (2n+1)\frac{T_\omega}{2}, \quad n = 0, 1, \ldots, \tag{29}$$

are observed in the distribution (27). Here $\omega$ and $T_\omega$ are the frequency and period of external periodic disturbance, respectively.

Moreover, analysis of Eq. (28) shows that the peaks (29) appear only in the presence of weak harmonic perturbation, in the contrary case monotone exponential decrease in the function (27) is observed. Thus, the physical interpretation of Eq. (27) is clear and consists in the following. An exponential background corresponds to that part of the residence-time distribution (27), which is generated by switchings caused by noise only, whereas the peaks (29) correspond to synchronization between the particle escape and external periodic perturbation.

In simulating stochastic resonance equation of Eq. (26) type in the framework of the bistable climate model just the use of optimal synchronization made is possible to obtain the "experimental" residence-time distribution function (Ganopolski and Rahmstorf, 2002; Braun et al, 2005). Moreover, the properties of simulated warm events, using which the residence-time distribution was obtained, was in good agreement with real properties of the Dansgaard-Oeschger events recorded by $\delta^{18}$O measurements of the Greenland ice (Andersen et al, 2004).

Comparing Eq. (25) corresponding to secular time scale and the bifurcation equations of global climate (1) and (2) constructed at millennial time scale, we can see that they have the identical structural components of ($ax-x^3+b$) type on right-hand side. As is shown in Ref. (Rusov et al, 2010), when the physical essence of process is correct "guessed" just the appearance of identical model structural components at different time scales is the key property of hierarchical models, which represents the principle of structural invariance of balance equations evolving at different time scales (Rusov et al, 2010). In other words, in our case this property lies in the fact that climate balance equations keep the self-similar macroscopic structure, for example, of the ($ax-x^3+b$) type and its governing parameters ($a$, $b$) in any models of the ECS regardless of used time scale.

In this sense, it is appropriate to remind the known model of oscillator with the "retarded argument" (Suarez and Schopf, 1988; Boutle et al, 2007) which quite truthfully describes a nonlinear dynamics of temperature oscillations (the $\Delta T$ anomalies) of the EL Nino current at annual time scale and in the modified form looks like:

$$\dot{x} = ax - x^3 + b - \mu \cdot x(t-\Delta) + \dot{Y}(t) + \Re + \xi(t), \quad x(t) \equiv \Delta T(t), \qquad (30)$$

where the first term on the right-hand side of Eq. (30) describes the strong positive feedback with the framework of the ocean-atmosphere system; the second term limits the growth of unstable perturbations of ocean temperature; $b$ is the constant parameter; the fourth term emulates the motion of equatorial "trapped" waves (the Rossby "latent" waves) through the Pacific ocean, which after elastic reflection on the western continental boundary return (as the Kelvin waves) along the equator to the central part of the Pacific ocean with the time delay $\Delta$; the fifth term Y($t$) describes the continuous changes of ocean surface average temperature in the course of year year; the sixth term ($\Re \approx 5\,°C/100$ years) emulates the global warming; $\xi(t)$ is the white noise of Gaussian type with zero mean ($\langle\xi(t)\rangle=0$) and standard deviation $\lambda$ (Suarez and Schopf, 1988; Boutle et al, 2007).

It is obvious, that the considered hieratic chain of ECS climatic states, i.e., the oscillation model of AL Nino temperature anomalies (at annual time scale) → the stochastic model of the Dansgaard-Oeschger heat anomalies during last glacial period (at secular time scale) → the bifurcation model of global climate (at millennial time scale), really contains the invariant macroscopic structure of ($ax-x^3+b$) type, that is a direct corollary of the principle of ECS balance equations structural invariance at different time scales. Moreover, it is necessary to add that the structural invariance is not only the direct indicator of correctly guessed physics of process at different time scales, but at the same time it establishes the obvious rules of time scale change.

Finally, we give short comment on the so-called "$CO_2$ doubling" problem. Analysis of the governing parameter $\tilde{b}(t)$ structure shows that it contains the three independent components (see Eq. (28) in Ref. (Rusov et al, 2010)). There are the insolation variation, the variation of reemission energy of water total mass (vapour and liquid) in the atmosphere and the climatic constant $\beta$, which determines the degree of variability of differential heat power reradiated by atmospheric carbon dioxide. At the same time, using Eq. (28) from Ref. (Rusov et al, 2010) it is easy to show that the following inequality

$$\frac{1}{2}\beta << \left| \eta_\alpha W_{reduced}(t) - 4\delta\sigma T_t^3 + \frac{1}{2}(2a_\mu T_t + b_\mu)H_\oplus(t) \right|, \quad for \quad \forall t \in [0, 730 kyr] \quad (31)$$

fulfils practically always at any $t=0 \div 740$ kyr (see Fig. 2).

The inequality (31) implies that an weighty influence of anthropogenic perturbation is possible only when the differential heat power reradiated by carbon dioxide $\beta_{perturb}$ is appreciably higher then the current value $\beta$ (see Eq.(14)). For example, performing the evident transformation of Eq. (28) from Ref. (Rusov et al, 2010), it is easy to show by computational experiment that in the time interval $t=0 \div 120$ kyr the threshold anthropogenic thermal effect appears at the hundred-fold increasing of the current value $\beta$. In other words, in the framework of the considered bifurcation model of the Earth global climate the so-called anthropogenic "$CO_2$ doubling" problem is practically absent.

Summarizing above-stated, we can conclude that the most important statements of the presented model is the fact that the Earth global climate, on the one hand, is completely described by the two governing parameters (insolation and galactic cosmic rays). On the other hand, if the theoretical or experimental time series, which describe an evolution of these governing parameters in the past or in the future, are available, the Earth global climate has not practically restrictions on temperature global forecast horizon at millennial time scale, or is quite predictable.